\documentclass{article}
\usepackage{spconf,amsmath,graphicx}

\usepackage{accents}
\usepackage{multirow}
\usepackage{xcolor}
\usepackage{wrapfig}

\DeclareMathOperator*{\argmax}{arg\,max}
\DeclareMathOperator*{\argmin}{arg\,min}
\newcommand\numberthis{\addtocounter{equation}{1}\tag{\theequation}}

\title{Full-Sum Decoding for Hybrid HMM based Speech Recognition\\ using LSTM Language Model}
\name{Wei Zhou, Ralf Schl\"uter, Hermann Ney}
\address{
Human Language Technology and Pattern Recognition, Computer Science Department,\\
  RWTH Aachen University, 52074 Aachen, Germany \\
AppTek GmbH, 52062 Aachen, Germany}

\begin{document}
%
\maketitle
\begin{abstract}
In hybrid HMM based speech recognition, LSTM language models have been widely applied and achieved large improvements. The theoretical capability of modeling any unlimited context suggests that no recombination should be applied in decoding. This motivates to reconsider full summation over the HMM-state sequences instead of Viterbi approximation in decoding. We explore the potential gain from more accurate probabilities in terms of decision making and apply the full-sum decoding with a modified prefix-tree search framework. The proposed full-sum decoding is evaluated on both Switchboard and Librispeech corpora. Different models using CE and sMBR training criteria are used. Additionally, both MAP and confusion network decoding as approximated variants of general Bayes decision rule are evaluated. Consistent improvements over strong baselines are achieved in almost all cases without extra cost. We also discuss tuning effort, efficiency and some limitations of full-sum decoding.
\end{abstract}
\section{Introduction \& Related Work}
\label{sec:intro}
Language models (LM) based on long short-term memory (LSTM) neural networks \cite{Hochreiter1997lstm} have been widely applied to automatic speech recognition. Both in lattice rescoring \cite{Sundermeyer2014latres, Kumar2017latres} and direct one-pass recognition \cite{Beck20191pass, Luescher2019libsp, Kitza2019swb}, large improvements over the $n$-gram count-based language models are observed. The recurrent update of cell states on each new word allows theoretically a capability of modeling any unlimited context.

One of the most common neural-network-based acoustic modeling methods is the hybrid hidden Markov model (HMM) approach \cite{Bourlard1993hybridhmm}, which still gives state-of-the-art performance on several widely used corpora \cite{Luescher2019libsp, Kitza2019swb}. The joint probability of a word sequence and input features requires a full summation over all HMM-state sequences. This leads to a high search complexity, since no exact word boundaries are present and all word sequences have to be tracked separately. However, with $n$-gram LM, the LM probability for paths sharing the same $n$-gram context is anyway the same. A complexity reduction based on this property is required. Viterbi approximation instead of full-sum is then widely applied, where the selection of the best word sequence hypothesis is only derived from its single best HMM state sequence. This allows an independent treatment on each state path of each word sequence. Additionally, exact word boundaries of each path are revealed. Recombination based on the same $n$-gram context can then be applied to largely simplify the search. 

But due to the unlimited context with LSTM LM, search is forced to again track each word sequence separately even in Viterbi decoding. Although recombination can be forced by taking a truncated context, certain performance degradation is observed \cite{Beck20191pass, Jorge20191pass}. This gives the motivation to reconsider the full-sum over HMM state sequences, where potential improvements of using better probability in decision making can be explored. 

Full-sum-based training criteria, such as connectionist temporal classification (CTC) \cite{Graves2006ctc}, have been widely adopted to improve model training. But even for models trained with full-sum related criteria, Viterbi decoding is still one standard approach for evaluation \cite{Kanda2016ctcmap, Willi2019sMBR}. One related work can be the prefix search decoding proposed in \cite{Graves2006ctc}, where certain heuristics are used to apply a section-wise full-sum search. This is claimed to outperform the best path decoding. \cite{Drexler2019subwordctc} extended it to subword level without the heuristics. Overall, there is only limited work to investigate full-sum-based decoding and its effect on decision making.

In this work, we revisit the joint probability used in decoding of hybrid HMM-based speech recognition. We argue that by applying full-sum instead of Viterbi approximation, improved probabilities should also benefit decision making. We apply the full-sum decoding using prefix-tree search with some modification. The proposed full-sum decoding is evaluated on both Switchboard and Librispeech corpora using state-of-the-art systems. Different models using cross-entropy (CE) and lattice-based state-level minimum Bayes risk (sMBR) \cite{Gibson2006MBR} training criteria are used. Additionally, both maximum a posteriori (MAP) and confusion network (CN) decoding as approximated variants of general Bayes decision rule are evaluated. Consistent improvements over the strong baselines are achieved in almost all cases. We also discuss tuning effort, efficiency and some limitations of the full-sum decoding.

\section{Full-sum Decoding}
\label{sec:fullsum}
Let $w_{1}^{N}$ and $x_{1}^{T}$ denote a word sequence of length $N$ and input features of $T$ frames, respectively. In the hybrid HMM approach, the key quantity used for decision making is their joint probability:
\begin{align*}
p(w_{1}^{N},x_{1}^{T}) & = p(w_{1}^{N}) \cdot \sum_{s_{1}^{T}} p(s_{1}^{T},x_{1}^{T}|w_{1}^{N})\\
                       & = p(w_{1}^{N}) \cdot \sum_{s_{1}^{T}} \prod_{t=1}^T p(x_t, s_t|s_{t-1})\\
                       & = p(w_{1}^{N}) \cdot \sum_{s_{1}^{T}} \prod_{t=1}^T \dfrac{p(s_t|x_t)}{p(s_t)} \cdot p(s_t|s_{t-1}) \numberthis \label{eqFS} \\
                       & \underset{\text{(Viterbi approximation)}}{= p(w_{1}^{N}) \cdot \max_{s_{1}^{T}} \prod_{t=1}^T \dfrac{p(s_t|x_t)}{p(s_t)} \cdot p(s_t|s_{t-1})} \numberthis \label{eqViterbi}
\end{align*}                    
where $s_{1}^{T}$ represents HMM-state sequences modeling the word sequence $w_{1}^{N}$. By applying Viterbi approximation, this joint probability for each word sequence is quantified by its single best state sequence only. Except for the numerical loss, the effect of such approximation in terms of final decision on best word sequence is unclear.

\subsection{Decision rules}
\label{sssec:decision}
In speech recognition, the general Bayes decision rule is given by:
\begin{align*}
x_{1}^{T} \rightarrow \accentset{\ast}{w}^N_1 = \argmin_{w_{1}^{N}} \sum_{\hat{w}_1^M} p(\hat{w}_1^M|x_{1}^{T}) \cdot \mathcal{L}(w_{1}^{N},\hat{w}_1^M)
\end{align*}
where $\mathcal{L}$ is the cost function between two word sequences. Instead of Levenshtein distance, the sentence-level 0-1 cost function is often used to simplify the optimization. This leads to the MAP decision rule:
\begin{align*}
x_{1}^{T} \rightarrow \accentset{\ast}{w}^N_1 = \argmax_{w_1^N} p(w_{1}^{N}|x_{1}^{T}) = \argmax_{w_1^N} p(w_{1}^{N},x_{1}^{T})
\end{align*}
Here differences of using Equation~\ref{eqFS} and Equation~\ref{eqViterbi} may result in both better and worse recognition.

The MAP decision rule aims to minimize the expected sentence error rate, which is suboptimal for speech recognition. A thorough study on different cost functions in terms of word error rate (WER) is given in \cite{Schluter2011br, Schluter2012cost}, but experimental verification is done using Equation~\ref{eqViterbi}. Decision rules other than MAP usually include further summations (e.g. for word posteriors), which can have possible interaction with the summation on HMM-level. Thus, the impact of full-sum on the decision becomes even more complicated. 

We also investigate this with CN decoding, which is a common approach to better approximate the Bayes decision rule. We follow the pivot-arc-based approach in \cite{hoffmeister11phd} to construct the CN by clustering the word arcs in the lattice into slots. The resulting hypotheses space is a superset of the original one and all paths have the same length. In this case, Hamming distance can be directly used as the cost function and the optimization problem is converted to slot-wise local decisions:
\begin{align*}
x_{1}^{T} \rightarrow \accentset{\ast}{w}_n = \argmax_{w_n} \sum_{\hat{w}_1^N:\hat{w}_n=w_n} p(\hat{w}_1^N|x_{1}^{T})
\end{align*}
Here the posterior probability is obtained by normalizing the joint probability over all word sequences in the word lattice.

\subsection{Prefix-tree search}
\label{ssec:dp}
For large vocabulary continuous speech recognition (LVCSR), explicit enumeration of all possible word sequences is computationally infeasible. Prefix-tree search using dynamic programming (DP) \cite{Ney1999dp} is a very efficient procedure to handle this. We also apply full-sum decoding using this approach with some modification. Without recombination, all state paths are characterized by their complete word history and probability summation within each word sequence becomes feasible. This is mainly realized by modifying the auxiliary $Q$ function in DP recursion:
\begin{align*}
Q_{\sum}(t,s;w_1^n) = \sum_{s'} \left\lbrace p(x_t, s| s';w_1^n) \cdot Q(t-1, s';w_1^n) \right\rbrace
\end{align*}
At each target state of each time frame, the probability summation over all incoming paths is carried out and this merged path continues further in the search. Paths of pronunciation variants are normalized and summed up as well. Note that without recombination, this does not increase complexity comparing to Viterbi decoding, since the major difference is just replacing the maximization with summation. The rest of the search procedure is rather straightforward. 

Standard score-based beam pruning is applied to maintain a reasonable size of search space. Ideally an infinite pruning threshold is needed to include all state paths of each word sequence, but we observe that the summation saturates at a normal threshold already. By increasing the pruning threshold, the number of paths grow dramatically. But the contribution of those paths with very low probability is negligible, and eventually has no effect on the results. For each final word sequence hypothesis, the joint probability in Equation~\ref{eqFS} contains the full-sum over all plausible state sequences.

Note that the resulting lattice from the full-sum decoding is a tree-like structure with certain approximation. Within each word sequence, word boundaries of each arc are taken from the single best state path. We use score to represent probabilities by taking $-log(.)$, which is common in decoding. The score at each arc boundary is computed from the probability summation of all partial paths merged at this boundary. The score marked on each word arc is the score difference between its right and left boundary, which does not correspond to the probability of this word. However, the accumulated score from the first till the last arc corresponds to the correct probability summation of all state paths for this word sequence. During CN construction, word boundaries are used as part of the arc distance measure for clustering and discarded afterwards. The correct full-sum is used to compute the posteriors for arc clustering and decision making.

\subsection{Parameter tuning}
\label{ssec:tuning}
In practice, acoustic scale $\alpha$ and LM scale $\beta$ are often applied for an optimal weighting among the knowledge sources. For MAP Viterbi decoding, $\alpha$ can be set to 1 and only $\beta$ needs to be tuned. But for full-sum decoding, both of them need to be tuned. This great tuning effort can be largely reduced by firstly tuning $\beta$ with $\alpha$ set to 1, and then linearly scaling both of them with the fixed ratio. To our experience, the same optimum is achieved in most cases as obtained by grid search.

\section{Experiments}
\label{sec:exp}

\subsection{Experimental settings}
\label{sssec:expsettings}
The proposed full-sum decoding is implemented based on the RWTH ASR toolkit \cite{rasr} with extensions described in \cite{Beck20191pass}. Both MAP and CN as approximated variants of Bayes decision rule are evaluated. For the MAP results, a one-pass recognition setup using LSTM LM without recombination is applied. Word lattices generated from the one-pass recognition are used to further apply CN decoding. Additionally, different models trained with CE and lattice-based sMBR criteria are evaluated. Note that full-sum is not applied in the training. Experiments are conducted on both the Switchboard corpus \cite{swb} and the Librispeech corpus \cite{libsp}. Acoustic and LM scales are optimized on the development sets.

\subsection{Switchboard}
\label{sssec:swb}
The acoustic models are the same as described in \cite{Willi2019sMBR, Kitza2019swb}, which consists of six BLSTM layers with 500 units for each direction. Both the CE and sMBR based models are trained on 300h Switchboard-1 Release 2 using 40-dimensional gammatone features \cite{Schluter2007gt}. The language model consists of two LSTM layers with 1024 units, which is described in \cite{Beck20191pass}. The Hub5'00 and Hub5'01 datasets are used as development and test set, respectively. 

Table~\ref{tab:swbwer} compares the proposed full-sum decoding to the standard Viterbi decoding with different models and decision rules. Only one case is improved with MAP decision rule, but consistent improvements are obtained with CN for both models. Interestingly for the CE model, the performance of Viterbi decoding even degrades from MAP to CN, which is not the case for full-sum decoding.

\begin{figure}[tb]
\begin{minipage}[b]{1.0\linewidth}
  \centering
  \centerline{\includegraphics[width=8.5cm]{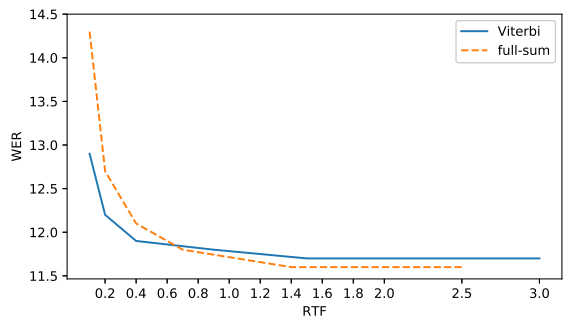}}
\end{minipage}
\caption{\it WER vs RTF comparison between Viterbi and full-sum decoding using the sMBR model and MAP decision rule on Hub5'00 dataset. (LSTM LM scoring is executed on a Nvidia Geforce 1080Ti GPU, and the rest of decoding is done on an Intel Xeon CPU E5-2620v4@2.1GHz.)}
\label{fig:swbwerrtf}
\end{figure}

Figure~\ref{fig:swbwerrtf} shows efficiency comparison of the two decodings with the sMBR model and MAP decision rule on the Hub5'00 dataset. Smaller real time factor (RTF) is obtained with stronger pruning. Full-sum decoding is more sensitive to strong pruning. This is because, besides direct search errors made by the strong pruning, it also suffers from an indirect influence. During search with strong pruning, non-negligible state paths of a partial word sequence hypothesis get pruned away, which introduces certain loss to the probability summation. This effect can accumulate across the search and eventually leads to more search errors. On the other hand, when WER converges with less pruning, there is no efficiency loss with the full-sum decoding.

\begin{table}
\caption{\it WER (in \%) comparison of full-sum and Viterbi decoding with different models and decision rules on the Hub5'00 and Hub5'01 datasets}
\begin{center}
\label{tab:swbwer}
\begin{tabular}{l|l|l|c|c}
\hline
Model                 & Decision            & Decoding & Hub5'00 & Hub5'01 \\ \hline
\multirow{4}{*}{CE}   & \multirow{2}{*}{MAP}     & Viterbi  & 12.2 & 12.2 \\ \cline{3-5}
                      &                          & full-sum & 12.2 & 12.2 \\ \cline{2-5}
                      & \multirow{2}{*}{CN} & Viterbi  & 12.4 & 12.4 \\ \cline{3-5}
                      &                          & full-sum & 12.1 & 12.2 \\ \hline
\multirow{4}{*}{sMBR} & \multirow{2}{*}{MAP}     & Viterbi  & 11.7 & 11.5 \\ \cline{3-5}
                      &                          & full-sum & 11.6 & 11.5 \\ \cline{2-5}
                      & \multirow{2}{*}{CN} & Viterbi  & 11.7 & 11.5 \\ \cline{3-5}
                      &                          & full-sum & 11.4 & 11.3 \\ \hline                                               
\end{tabular}
\end{center}
\end{table}

\begin{table*}
\centering
\caption{\it WER (in \%) comparison of full-sum and Viterbi decoding with different models and decision rules on the Librispeech datasets}
\label{tab:libspwer}
\begin{tabular}{l|l|l|c|c|c|c}
\hline
Model                 & Decision & Decoding & dev-clean & dev-other & test-clean & test-other \\ \hline
\multirow{4}{*}{CE}   & \multirow{2}{*}{MAP}     & Viterbi    & 2.4 & 5.8 & 2.8 & 6.3 \\ \cline{3-7}
                      &                          & full-sum   & 2.4 & 5.7 & 2.8 & 6.2 \\ \cline{2-7}
                      & \multirow{2}{*}{CN} & Viterbi    & 2.4 & 5.7 & 2.8 & 6.1 \\ \cline{3-7}
                      &                          & full-sum   & 2.4 & 5.6 & 2.8 & 6.0 \\ \hline
\multirow{4}{*}{sMBR} & \multirow{2}{*}{MAP}     & Viterbi    & 2.2 & 5.1 & 2.6 & 5.5 \\ \cline{3-7}
                      &                          & full-sum   & 2.1 & 5.0 & 2.5 & 5.4 \\ \cline{2-7}
                      & \multirow{2}{*}{CN} & Viterbi    & 2.2 & 5.0 & 2.6 & 5.4 \\ \cline{3-7}
                      &                          & full-sum   & 2.1 & 4.8 & 2.5 & 5.3 \\ \hline                           
\end{tabular}
\end{table*}

\begin{figure}[tb]
\begin{minipage}[b]{1.0\linewidth}
  \centering
  \centerline{\includegraphics[width=8.5cm]{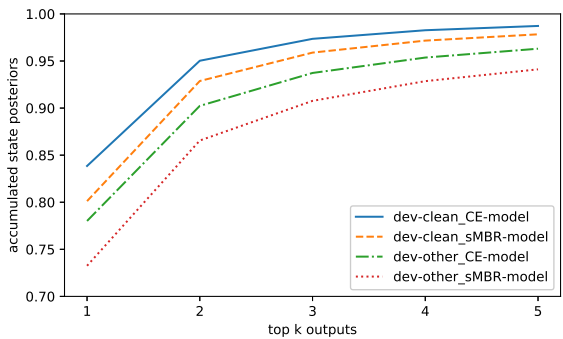}}
\end{minipage}
\caption{\it Average accumulated state posteriors of models' top 1-5 outputs at each frame on the Librispeech dev datasets. CE and sMBR denote the training criteria of the two models.}
\label{fig:libspposterior}
\end{figure}

\subsection{Librispeech}
\label{sssec:libsp}
We use the state-of-the-art hybrid HMM system described in \cite{Luescher2019libsp}. Both CE and sMBR based acoustic models consist of six BLSTM layers with 1000 units for each direction, and are trained on the complete Librispeech training data with 50-dimensional gammatone features. The language model has two LSTM layers with 4096 units.  

Detailed WER comparison of the two decodings with different models and decision rules are shown in Table~\ref{tab:libspwer}. With such a strong baseline of very low WER, further improvements are obtained with the full-sum decoding in almost all scenarios. This again verifies the benefit of improved probabilities in terms of decision making. Note that there is no difference in the models but just a different way to apply them in decoding, therefore the improvements come without additional cost at training.

One interesting observation is that more improvements are achieved on the 'other' datasets than the 'clean' ones. To better understand this, we further check the state posteriors from acoustic models' top 5 outputs at each frame. We accumulate these posteriors at each frame and average them within each dev set for each model, which is shown in Figure~\ref{fig:libspposterior}. This partially reflects the models' distribution over the states and both models tend to produce sharper distribution on the dev-clean set than on the dev-other set. One way to interpret this is that when the models are more confident on a very limited number of state paths, there is likely less difference between full-sum and Viterbi decoding. 

\section{Conclusion}
\label{sec:con}
In this paper, we showed that by applying full-sum instead of Viterbi approximation in decoding, more accurate probabilities improve decision making. The proposed full-sum decoding was verified on different corpora, models and decision rules, and showed consistent improvements in almost all cases. We showed that the full-sum decoding is more sensitive to strong pruning, but has no efficiency loss with normal pruning. One major advantage is that by applying the proposed full-sum decoding, even for the state-of-the-art system with very low WER, further improvements can be achieved with negligible extra cost. 

One possible future direction is to investigate the effect of full-sum decoding in noisy conditions where models are in general less confident. Another direction is to investigate the joint effect on models trained with full-sum-based training criteria, such as CTC.

\section{Acknowledgements}
\vspace{-2mm}
\begin{wrapfigure}[5]{l}{0.13\textwidth}
	\vspace{-4mm}
	\begin{center}
		\includegraphics[width=0.15\textwidth]{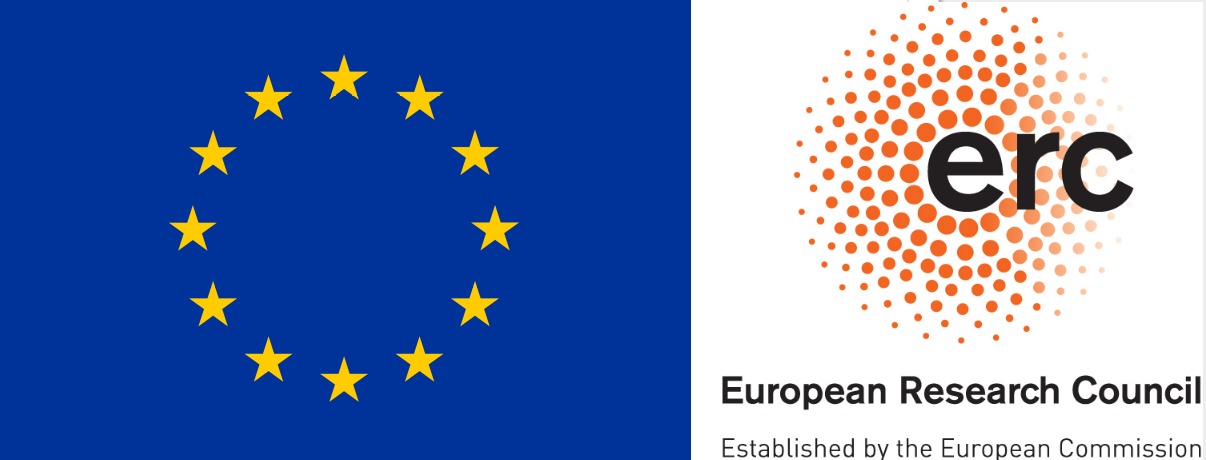} \\
	\end{center}
	\vspace{-4mm}
\end{wrapfigure}
\small
This work has received funding from the European Research Council (ERC) under the European Union's Horizon 2020 research and innovation program (grant agreement No 694537, project ``SEQCLAS") and from a Google Focused Award. The work reflects only the authors' views and none of the funding parties is responsible for any use that may be made of the information it contains.

We thank Christoph L\"{u}scher, Kazuki Irie, Markus Kitza and Wilfried Michel for providing the models.

\bibliographystyle{IEEEbib}
\bibliography{refs}

\end{document}